\documentclass{aa}
\usepackage[varg]{txfonts}
\usepackage[colorlinks=true,citecolor=blue,linkcolor=blue]{hyperref}
\usepackage{changes}

\begin{document}

\title{On the nature of some SGRs and AXPs as rotation-powered neutron stars}

\author{Jaziel G.~Coelho\inst{1,2}
\thanks{\emph{jaziel.coelho@inpe.br}}
\and 
D. L. C\'aceres\inst{2}
\and
R. C. R. de Lima\inst{3,4}
\and
M. Malheiro\inst{5} 
\and 
J. A. Rueda\inst{2,3,6}
\thanks{\emph{jorge.rueda@icra.it}}
\and 
R. Ruffini\inst{2,3,6}
\thanks{\emph{ruffini@icra.it}}
}

\institute{
Div. Astrof\'isica, Instituto Nacional de Pesquisas Espaciais, Av. dos Astronautas 1758, 12227--010 S\~ao Jos\'e dos Campos, Brazil 
\and
Dipartimento di Fisica and ICRA, Sapienza Universit\`a di Roma, P.le Aldo Moro 5, I--00185 Rome, Italy
\and
ICRANet, P.zza della Repubblica 10, I--65122 Pescara, Italy 
\and 
Universidade do Estado de Santa Catarina, Av. Madre Benvenuta, 2007 Itacorubi, 88.035--901, Florian\'opolis, Brazil
\and
Instituto Tecnol\'ogico de Aeron\'autica, Pra\c{c}a Marechal Eduardo Gomes, 50 Vila das Ac\'acias, 12228--900 S\~ao Jos\'e dos Campos, Brazil 
\and 
ICRANet-Rio, Centro Brasileiro de Pesquisas F\'isicas, Rua Dr. Xavier Sigaud 150, 2290--180 Rio de Janeiro, Brazil}

\date{Accepted XXX. Received YYY; in original form ZZZ}

\abstract{
{\it Context.} Soft gamma repeaters (SGRs) and anomalous X-ray pulsars (AXPs) are slow rotating isolated pulsars whose energy reservoir is still matter of debate. Adopting neutron star (NS) fiducial parameters; mass $M=1.4M_\odot$, radius $R=10$~km, and moment of inertia, $I=10^{45}$~g~cm$^2$, the rotational energy loss, $\dot{E}_{\rm rot}$, is lower than the observed luminosity (dominated by the X-rays) $L_X$ for many of the sources.\\
{\it Aims.} We investigate the possibility that some members of this family could be canonical rotation-powered pulsars using realistic NS structure parameters instead of fiducial values.\\
{\it Methods.} We compute the NS mass, radius, moment of inertia and angular momentum from numerical integration of the axisymmetric general relativistic equations of equilibrium. We then compute the entire range of allowed values of the rotational energy loss, $\dot{E}_{\rm rot}$, for the observed values of rotation period $P$ and spin-down rate $\dot{P}$. We also estimate the surface magnetic field using a general relativistic model of a rotating magnetic dipole.\\
{\it Results.} We show that realistic NS parameters lowers the estimated value of the magnetic field and radiation efficiency, $L_X/\dot{E}_{\rm rot}$, with respect to estimates based on fiducial NS parameters. We show that nine SGRs/AXPs can be described as canonical pulsars driven by the NS rotational energy, for $L_X$ computed in the soft (2--10~keV) X-ray band. We compute the range of NS masses for which $L_X/\dot{E}_{\rm rot}<1$. We discuss the observed hard X-ray emission in three sources of the group of nine potentially rotation-powered NSs. This additional hard X-ray component dominates over the soft one leading to $L_X/\dot{E}_{\rm rot}>1$ in two of them.\\
{\it Conclusions.} We show that 9 SGRs/AXPs can be rotation-powered NSs if we analyze their X-ray luminosity in the soft 2--10~keV band. Interestingly, four of them show radio emission and six have been associated with supernova remnants (including Swift J1834.9-0846 the first SGR observed with a surrounding wind nebula). These observations give additional support to our results of a natural explanation of these sources in terms of ordinary pulsars. Including the hard X-ray emission observed in three sources of the group of potential rotation-powered NSs, this number of sources with $L_X/\dot{E}_{\rm rot}<1$ becomes seven. It remains open to verification 1) the accuracy of the estimated distances and 2) the possible contribution of the associated supernova remnants to the hard X-ray emission.
}
\maketitle
\section{Introduction}\label{sec:1}

SGRs and AXPs constitute a class of pulsars with the following main properties \citep{2008A&ARv..15..225M}: rotation periods $P\sim(2$--$12)$~s, slowing down rates $\dot{P}\sim(10^{-15}-10^{-10})$~s/s, persistent X-ray luminosity as large as $10^{35}$ erg~s$^{-1}$ and transient activity in the form of outbursts of energies around $(10^{41}$--$10^{43})$~erg. Giant flares of even larger energies, $(10^{44}$--$10^{47})$~erg, up to now only observed in SGRs.

A spinning down NS loses rotational energy at a rate given by
\begin{equation}\label{eq:EdotNS}
\dot{E}_{\rm rot}=-4 \pi^2 I \frac{\dot{P}}{P^3}
\end{equation}
which, adopting fiducial moment of inertia $I = 10^{45}$~g~cm$^2$, becomes
\begin{equation}\label{eq:EdotNSfid}
\dot{E}^{\rm fid}_{\rm rot}=-3.95\times 10^{46} \frac{\dot{P}}{P^3}\quad {\rm erg~s}^{-1}.
\end{equation}
Correspondingly to the fiducial moment of inertia, usual fiducial values for the mass and radius of a NS adopted in the literature are, respectively, $M=1.4 M_\odot$ and radius $R=10$~km. For the observed values of $P$ and $\dot{P}$, Eq.~(\ref{eq:EdotNSfid}) leads to values lower than the observed X-ray luminosity from SGR/AXPs (i.e.~$\dot{E}^{\rm fid}_{\rm rot}<L_X$). This is in contrast with traditional rotation-powered pulsars which show $\dot{E}_{\rm rot} > L_{X}^{\rm obs}$. 

The apparent failure of the traditional energy reservoir of pulsars in SGR/AXPs has led to different scenarios for the explanation of SGRs and AXPs, e.g.: magnetars~\citep{1992ApJ...392L...9D,1995MNRAS.275..255T}; drift waves near the light-cylinder of NSs \citep[see][and references therein]{2010ARep...54..925M}; fallback accretion onto NSs \citep{2013ApJ...764...49T}; accretion onto exotic compact stars such as quark stars \citep{2006MNRAS.373L..85X}; the quark-nova remnant model \citep{2011MNRAS.415.1590O}; and massive, fast rotating, highly magnetized WDs \citep{2012PASJ...64...56M,2013A&A...555A.151B,2013ApJ...772L..24R,2014PASJ...66...14C}

None of the above scenarios appears to be ruled out by the current observational data, thus further scrutiny of the nature and the possible energy reservoir of SGRs and AXPs deserves still attention.

Following the above reasoning, we aim to revisit in this work the possibility that some SGR/AXPs could be rotation-powered NSs, but now exploring the entire range of NS parameters allowed by the conditions of stability of the star, and not only on the use of fiducial parameters. We have already examined this possibility in \citet{2012PASJ...64...56M,2013arXiv1307.8158C,2014PASJ...66...14C,2015mgm..conf.2465C,doi:10.1142/S021827181641025X} and have found at the time four sources (1E 1547.0--5408, SGR 1627--41, PSR J1622--4950, and XTE J1810--197) of the SGR/AXPs catalog explainable as rotation-powered NSs \citep[see, also,][]{2012ApJ...748L..12R}. We show in this article that this conclusion can be indeed extended to other seven sources. For the total eleven objects we report the range of masses where the rotation-power condition $\dot{E}_{\rm rot} > L_X$ is satisfied.

Once identified as rotation-powered NSs, one is led to the theoretical prediction that some of the phenomena observed in ordinary pulsars could also be observed in SGRs and AXPs. Indeed, we found that for the above 11 SGRs/AXPs describable as rotation-powered NSs:
\begin{itemize}
\item
The energetics of their observed outbursts can be explained from the gain of rotational energy during an accompanied glitch. Such a glitch-outburst connection is not expected in a source not driven by rotational energy.

\item 
The radio emission, a property common in pulsars but generally absent in SGRs and AXPs, is observed in four of these objects \citep[see, e.g.,][]{2005ApJ...632L..29H,2006Natur.442..892C,2006HEAD....9.0603H,2007ApJ...663..497C,2007ApJ...666L..93C,  2010ApJ...721L..33L,2012MNRAS.422.2489L,2013Natur.501..391E}.

\item 
Six sources have possible associations with supernova remnants (SNRs), including Swift J1834.9-0846 the first SGR for which a pulsar wind nebula has been observed \citep{2016arXiv160406472Y}.
\end{itemize}

We also analyze the observed hard X-ray emission in the 20-150~keV band in 5 of the above 11 sources. As we shall discuss, this emission dominates over the soft X-rays leading to $L_X/\dot{E}_{\rm rot}>1$ in 4 of them. With this, the number of potential rotation-powered sources becomes seven. However, this conclusion remains open for further verification since it critically depends 1) on the accuracy of the estimated distances to the sources and 2) on the possible contribution of the supernova remnants present in the hard X-ray component.

This article is organized as follows. We first compute in section~\ref{sec:2} the structure properties of NSs, and then in section~\ref{sec:3} we estimate the surface magnetic field using both realistic structure parameters and a general relativistic model of a rotating magnetic dipole. We compute in section~\ref{sec:4} the ratio $L_X/\dot{E}_{\rm rot}$ for all the SGRs/AXPs for the entire range of possible NS masses. We also show in section~\ref{sec:5} an analysis of the glitch/outburst connection in the nine aforementioned sources. Finally, in Section~\ref{sec:7}, we summarize the main conclusions and remarks.

\section{Neutron star structure}\label{sec:2}

In order to compute the rotational energy loss of a NS as a function of its structure parameters, e.g. mass and radius, we need to construct the equilibrium configurations of a uniformly rotating NS in the range of the observed periods. We have shown in \citet{2011PhLB..701..667R,2011NuPhA.872..286R,2012NuPhA.883....1B,2014NuPhA.921...33B} that, in the case of both static and rotating NSs, the Tolman-Oppenheimer-Volkoff (TOV) system of equations \citep{1939PhRv...55..374O,1939PhRv...55..364T} is superseded by the Einstein-Maxwell system of equations coupled to the general relativistic Thomas-Fermi equations of equilibrium, giving rise to what we have called the Einstein-Maxwell-Thomas-Fermi (EMTF) equations. In the TOV-like approach, the condition of local charge neutrality is applied to each point of the configuration, while in the EMTF equations the condition of global charge neutrality is imposed. The EMTF equations account for the weak, strong, gravitational and electromagnetic interactions within the framework of general relativity and relativistic nuclear mean field theory. In this work we shall use both global (EMTF) and local (TOV) charge neutrality to compare and contrast their results.

\subsection{Nuclear EOS}

\begin{figure}
\centering
\includegraphics[width=\columnwidth,clip]{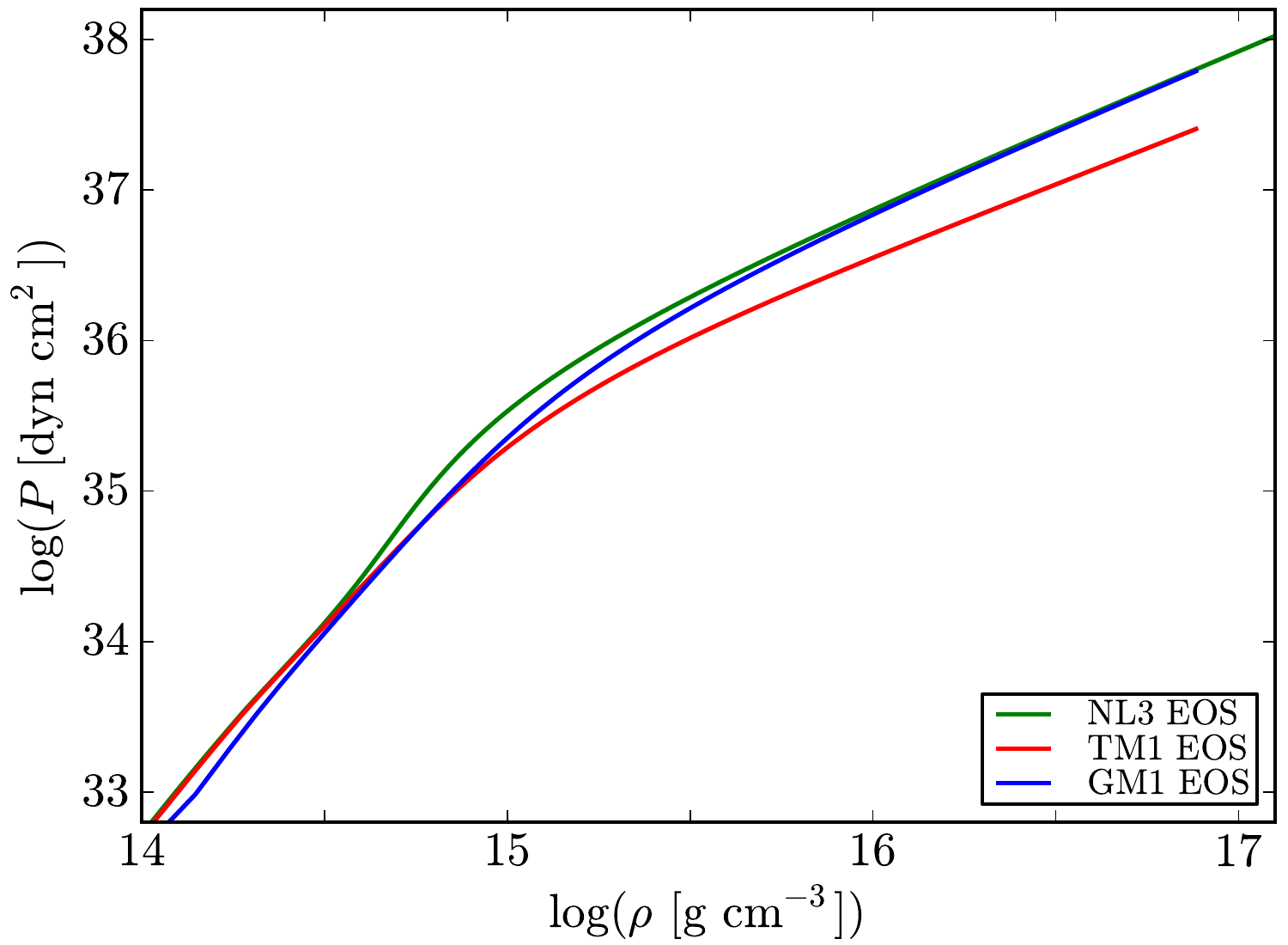}
\caption{NL3, TM1, and GM1 EOS behavior at sub and supranuclear densities.}\label{fig:EOS}
\end{figure}

The NS interior is made up of a core and a crust. The core of the star has densities higher than the nuclear one, $\rho_{\rm nuc}\approx 3\times 10^{14}$~g~cm$^{-3}$, and it is composed of a degenerate gas of baryons (e.g.~neutrons, protons, hyperons) and leptons (e.g.~electrons and muons). The crust, in its outer region ($\rho \leq \rho_{\rm drip}\approx 4.3\times 10^{11}$~g~cm$^{-3}$), is composed of ions and electrons, and in the inner crust ($\rho_{\rm drip}<\rho<\rho_{\rm nuc}$), there are also free neutrons that drip out from the nuclei. For the crust, we adopt the Baym-Pethick-Sutherland (BPS) EOS \citep{1971ApJ...170..299B}, which is based on the \citet{1971NuPhA.175..225B} work. For the core, we adopt relativistic mean-field (RMF) theory models. We use an extension of the \citet{1977NuPhA.292..413B} formulation with a massive scalar meson ($\sigma$) and two vector mesons ($\omega$ and $\rho$) mediators, and possible interactions amongst them. We adopt in this work three sets of parameterizations of these models (see table~\ref{tab1} and figure~\ref{fig:EOS}): the NL3 \citep{1997PhRvC..55..540L}, TM1 \citep{1994NuPhA.579..557S}, and GM1 \citep{1991PhRvL..67.2414G} EOS.
\begin{table}
\centering
\caption{Meson masses and coupling constants in the parameterizations NL3, TM1, and GM1}\label{tab1}
\begin{tabular}{l r r r r}
	\hline
		 & & NL3 & TM1 & GM1 \\
		 \hline
		 $M$(MeV) & & $939.00$ & $938.00$ & $938.93$ \\
		 $m_{\sigma}$(MeV)& & $508.194$ & $511.198$ & $512.000$\\
		 $m_{\omega}$(MeV)& & $782.501$ & $783.000$ & $783.000$ \\
		 $m_{\rho}$(MeV)& & $763.000$ & $770.000$ & $770.000$\\
		 $g_{\sigma}$ & & $10.2170$ & $10.0289$ & $8.9073$ \\
		 $g_{\omega}$ & & $12.8680$ & $12.6139$ & $10.6089$\\
		 $g_{\rho}$ & & $4.4740$ & $7.2325$ & $4.0972$\\
		 \hline
\end{tabular}
\end{table}

\subsection{Mass-Radius Relation and Moment of Inertia}

\begin{figure}
\centering
\includegraphics[width=\hsize,clip]{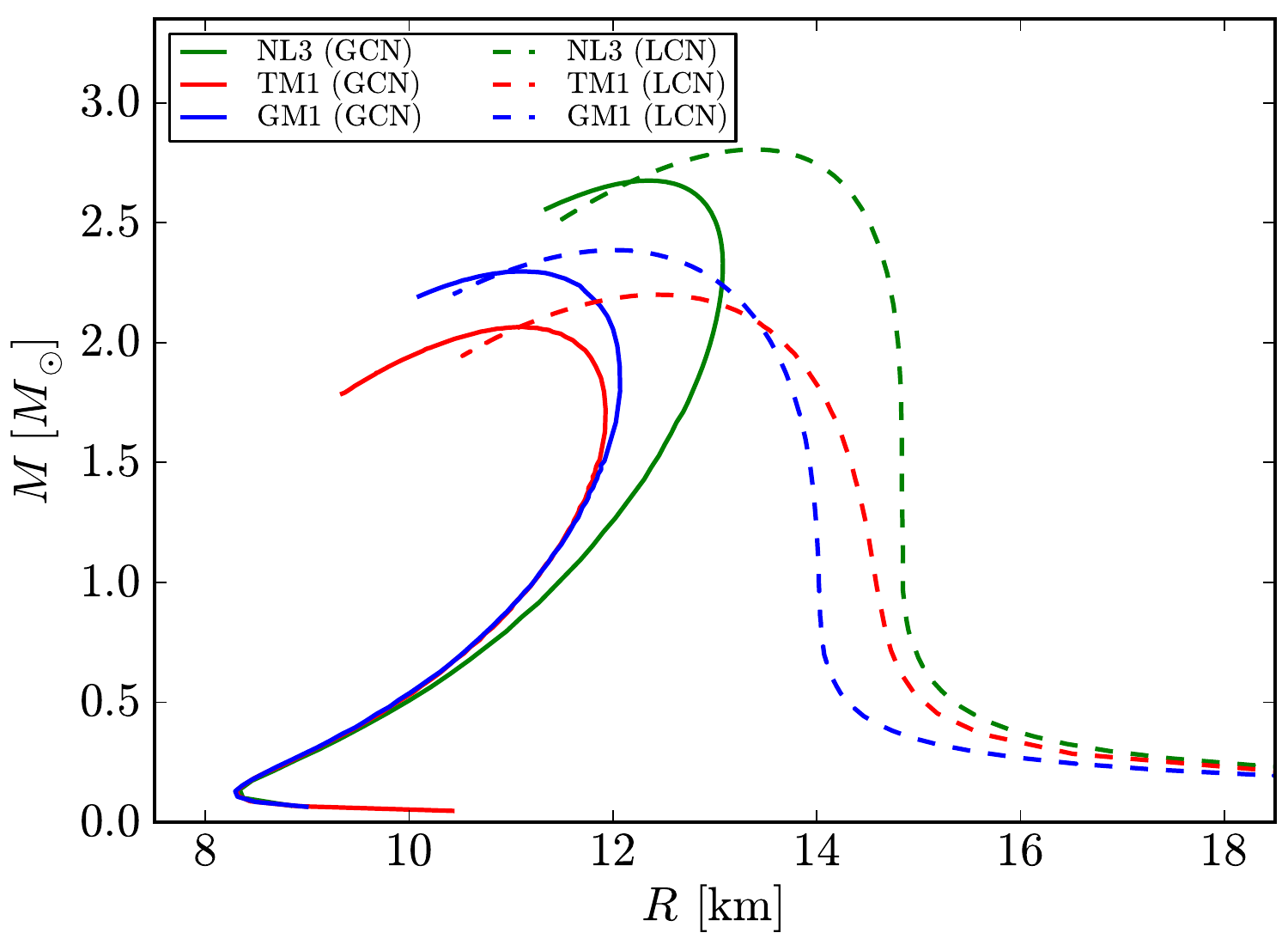}
\caption{Mass-Radius relation for the NL3, TM1, and GM1 EOS in the cases of global (solid curves) and local (dashed curves) charge neutrality.}\label{fig:MR}
\end{figure}

For the rotational periods as the ones observed in SGR/AXPs ($P\sim 2$--$12$~s), the structure of the rotating NS can be accurately described by small rotation perturbations from the spherically symmetric configuration \citep[see, e.g.,][]{2014NuPhA.921...33B,2015ApJ...799...23B}, 
using the Hartle's formalism \citep{1967ApJ...150.1005H}. Following this method we compute rotating configurations, accurate up to second-order in $\Omega$, with the same central density as the seed static non-rotating configurations. The mass-radius relation for non-rotating configurations in the cases of global and local charge neutrality are shown in Fig.~\ref{fig:MR}. For the rotation periods of interest here, the mass-equatorial radius relation of the uniformly rotating NSs practically overlaps the one given by the static sequence \citep[see Fig. 1 in][]{2015ApJ...799...23B}. Thus, we take here advantage of this result and consider hereafter, as masses and corresponding radii, the values of the non-rotating NSs.

The moment of inertia is given by
\begin{equation}
I = \frac{J}{\Omega},
\end{equation}
where $\Omega$ is the angular velocity and $J$ is the angular momentum given by
\begin{equation}
J = \frac{1}{6} R^4 \left( \frac{d\bar{\omega}}{dr} \right)_{r=R}.
\end{equation}
Here $R$ is the radius of the non-rotating star with the same central density as the rotating one, $\bar{\omega}=\Omega - \omega(r)$ is the angular velocity of the fluid relative to the local inertial frame, and $\omega$ is the  angular velocity of the local frame. The angular velocity $\Omega$ is related the angular momentum $J$ by
\begin{equation}
\Omega = \bar{\omega} (R) + \frac{2J}{R^3}.
\end{equation}

\begin{figure*}
\centering
\includegraphics[width=0.5\hsize,clip]{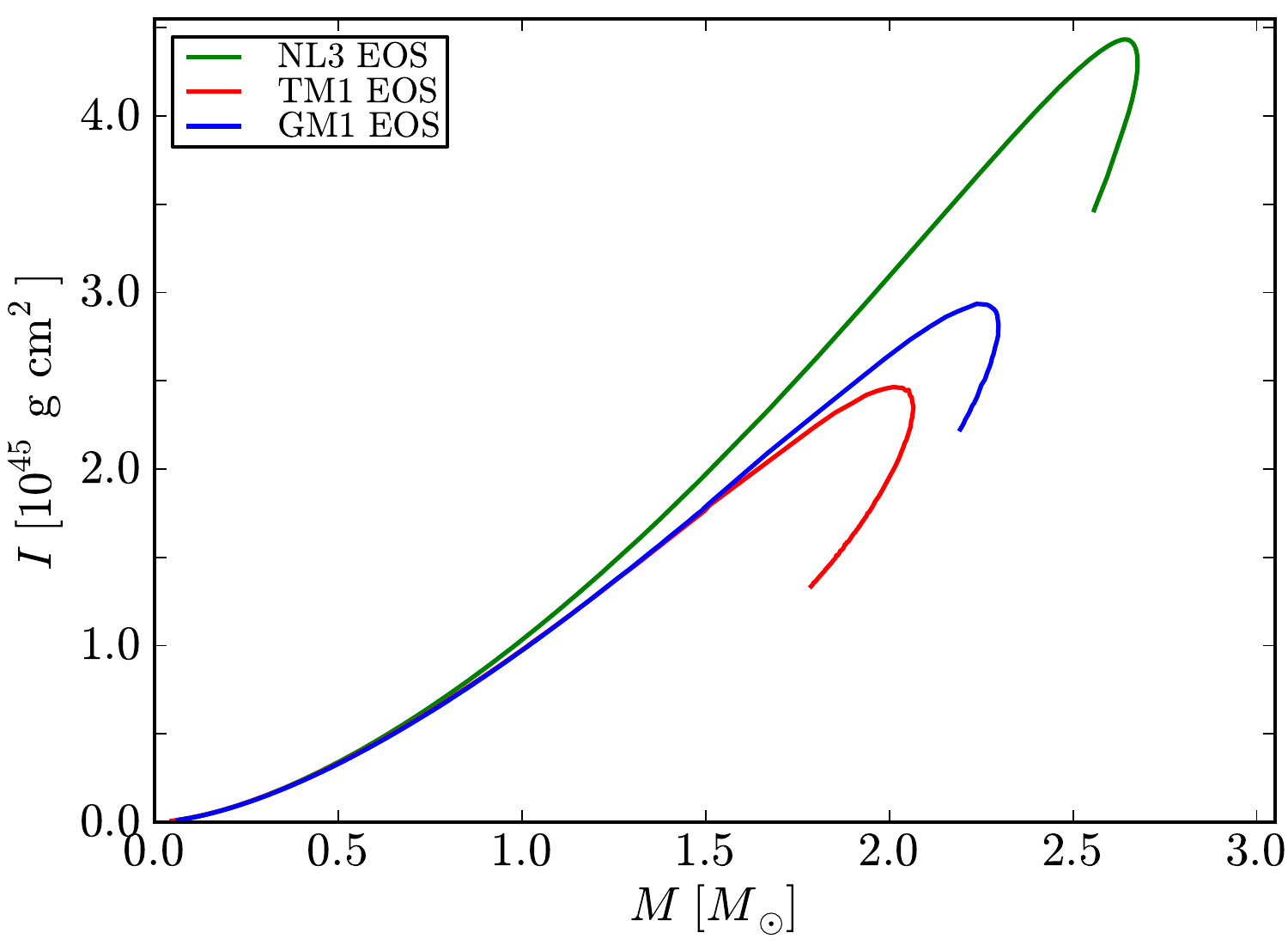}\includegraphics[width=0.5\hsize,clip]{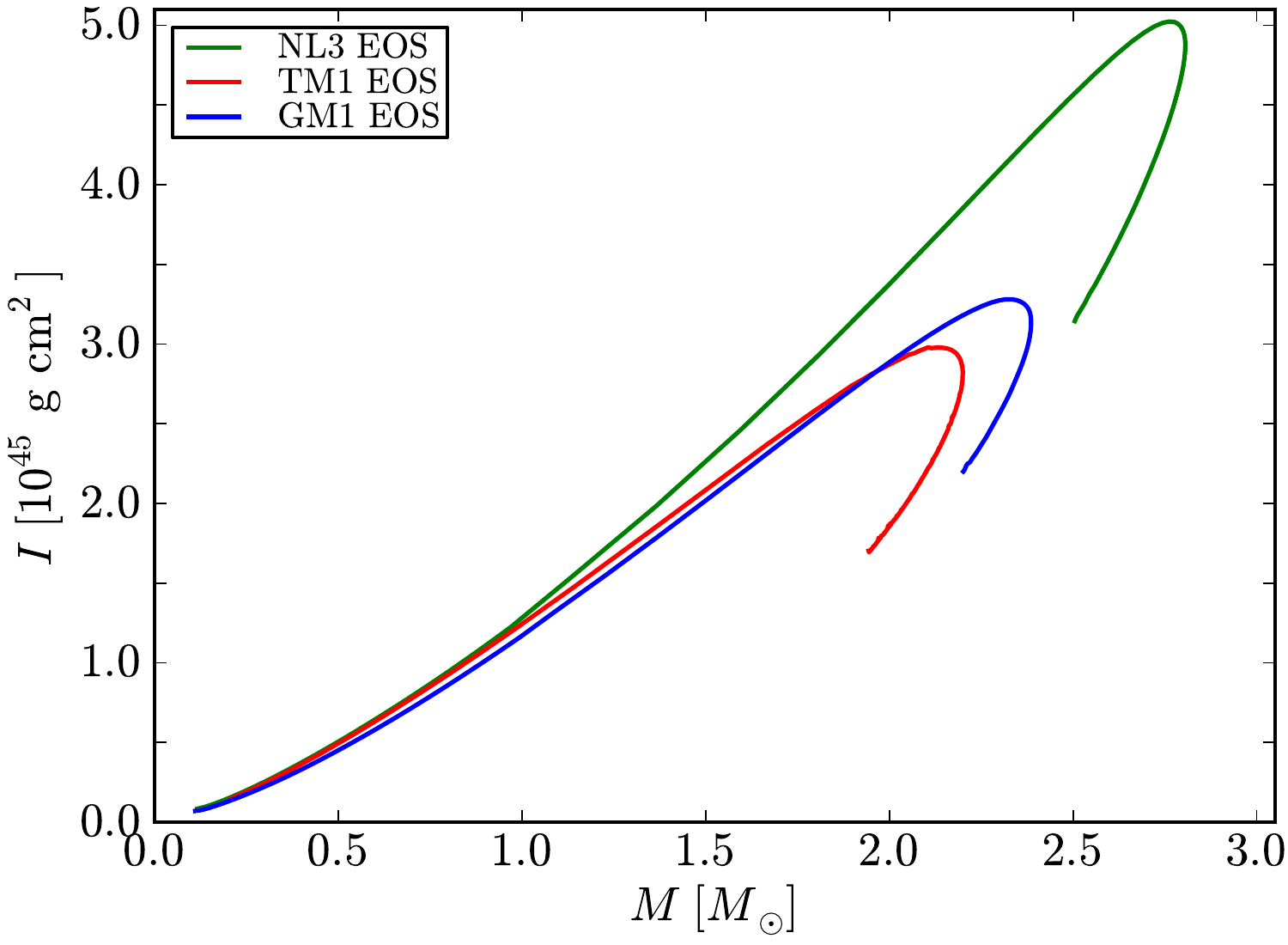}
\caption{Moment of inertia as a function of the NS mass for the NL3, TM1, and GM1 EOS in the cases of global (left panel) and local (right panel) charge neutrality.}\label{fig:IvsM}
\end{figure*}

Fig.~\ref{fig:IvsM} shows the behavior of the moment of inertia as a function of the mass of the NS for the three EOS NL3, TM1 and GM1 and both in the case of global and local charge neutrality. Although in general there is a dependence of all the structure parameters on the nuclear EOS, we use below, without loss of generality and for the sake of exemplification, only the GM1 EOS. Similar qualitatively and quantitatively results are obtained for the other EOSs. It is worth mentioning that the chosen EOS lead to a maximum stable mass larger than $2~M_\odot$, the heaviest NS mass measured \citep{2010Natur.467.1081D,2013Sci...340..448A}.

\section{Surface Magnetic Field}\label{sec:3}
Since the range of $P$ for SGRs and AXPs is similar to the one concerning the high-magnetic field pulsar class, we can directly apply the results of \citet{2015ApJ...799...23B}, applying only the most relevant correction for this range of periods to estimate the surface magnetic field, namely the finite-size correction. The exact solution of the radiation power of a (slowly) rotating, magnetic dipole, which duly generalizes the classic solution by \citet{1955AnAp...18....1D}, is given by \citep[see][and references therein]{2004MNRAS.352.1161R}
\begin{equation}\label{eq:dipoleGR}
P^{\rm GR}_{\rm dip}=-\frac{2}{3}\frac{\mu^2_\bot \Omega^4}{c^3} \left(\frac{f}{N^2}\right)^2\, ,
\end{equation}
where $\mu_\bot= \mu \sin \chi$, is the component of the magnetic dipole moment perpendicular to the rotation axis, $\mu = B R^3$ with $B$ the surface magnetic field at the star's equator, $\chi$ is the inclination angle between the magnetic dipole and rotation axis, and $f$ and $N$ are the general relativistic corrections
\begin{eqnarray}\label{eq:fN}
f &=&-\frac{3}{8}\left(\frac{R}{M}\right)^3\left[\ln(N^2)+\frac{2 M}{R}\left(1+\frac{M}{R}\right)\right]\, ,\\
N &=&\sqrt{1-\frac{2 M}{R}}\, ,
\end{eqnarray}
with $M$ the mass of the non-rotating configuration. Now, equating the rotational energy loss, Eq.~(\ref{eq:EdotNS}) to the above electromagnetic radiation power, Eq.~(\ref{eq:dipoleGR}), one obtains the formula to infer the surface magnetic field, given the rotation period and the spindown rate:
\begin{equation}\label{eq:BGR}
B_{\rm GR}=\frac{N^2}{f} \left(\frac{3 c^3}{8 \pi^2} \frac{I}{R^6} P \dot{P} \right)^{1/2},
\end{equation}
where we have introduced the subscript `GR' to indicate explicitly the magnetic field inferred from the above general relativistic expression, and we have adopted for simplicity an inclination angle $\chi=\pi/2$. 

\begin{figure}
\centering
\includegraphics[width=\columnwidth,clip]{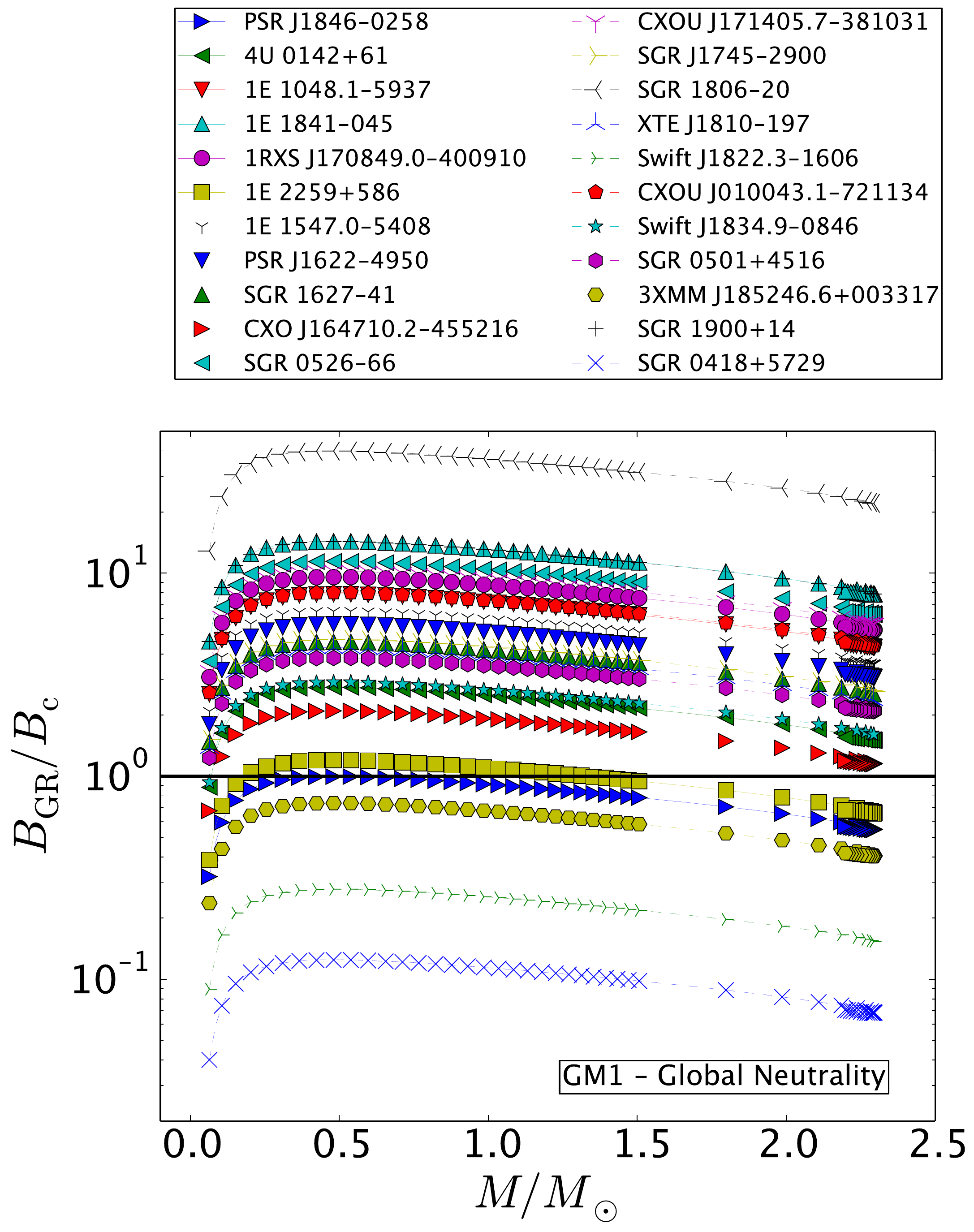}
\caption{(Color online) Magnetic field $B_{\rm GR}$ given by equation (\ref{eq:BGR}), in units of $B_c=m^2_e c^3/(e \hbar)=4.4\times 10^{13}$~G, as function of the mass (in solar masses) in the global charge neutrality case.}\label{figure1}
\end{figure}

\begin{figure}
\centering
\includegraphics[width=\columnwidth,clip]{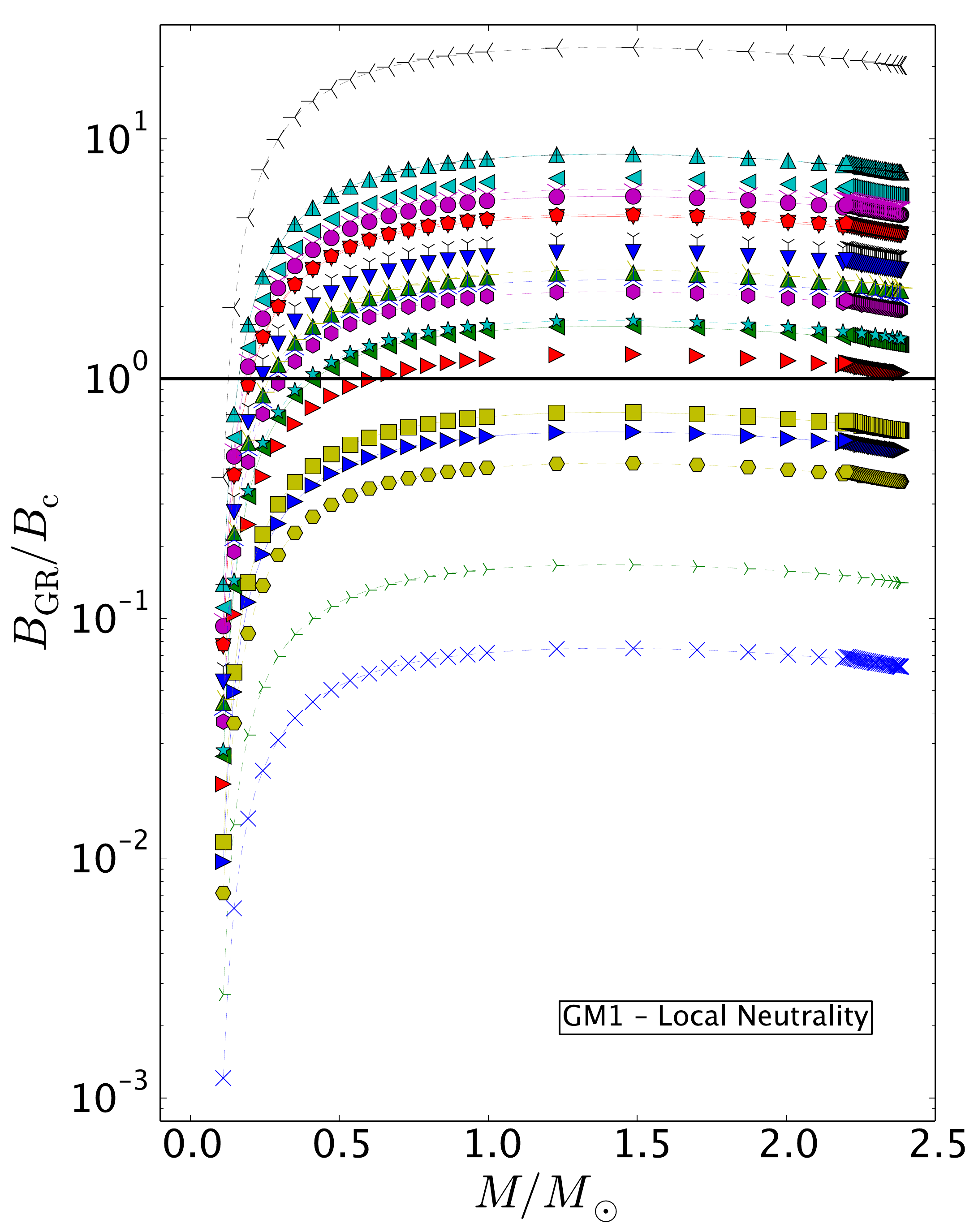}
\caption{(Color online) Magnetic field $B_{\rm GR}$ given by equation (\ref{eq:BGR}), in units of $B_c=4.4\times 10^{13}$~G, as function of the mass (in solar masses) in the local charge neutrality case.}\label{figure2}
\end{figure}

Figs.~\ref{figure1} and~\ref{figure2} show our theoretical prediction for the surface magnetic fields of the SGR/AXPs as a function of the NS mass, using Eq.~(\ref{eq:BGR}), for the GM1 EOS and for the global and local charge neutrality cases, respectively. We find that in both cases some of the sources have inferred magnetic fields lower than the critical value, $B_c$, for some range of NS masses. Clearly this set of sources includes SGR 0418+5729, Swift J1822.3-1606 and 3XMM J185246.6+003317, which are already known to show this feature even using fiducial NS parameters and the classic magnetic dipole model \citep[see e.g.,][]{2014ApJS..212....6O}. It is worth to note that Eq.~(\ref{eq:BGR}) is derived for a rotating magnetic dipole in electrovacuum, thus neglecting the extra torque from the presence of magnetospheric plasma. The addition of this torque certainly leads to values of the magnetic field still lower than the ones shown here. However, the inclusion of the torques from the magnetosphere is beyond the scope of this work and will be taken into account in future works.

\section{SGRs and AXPs Efficiency}\label{sec:4}

Another important quantity for the identification of the nature of the sources is the radiation efficiency, namely the ratio between the observed luminosity and the rotational energy loss (\ref{eq:EdotNS}). It is clear from figure~\ref{fig:IvsM} that such a ratio is a function of the NS mass, via the moment of inertia. For SGR/AXPs the dominant emission is in X-rays, thus we analyze all the possible values of the ratio $L_X/\dot{E}_{\rm rot}$ in the entire parameter space of NSs. As we show below, some SGRs and AXPs allow a wide range of masses for which $L_X/\dot{E}_{\rm rot} \lesssim 1$, implying a possible rotation-powered nature for those sources.

\begin{figure*}
\centering
\includegraphics[width=0.5\hsize,clip]{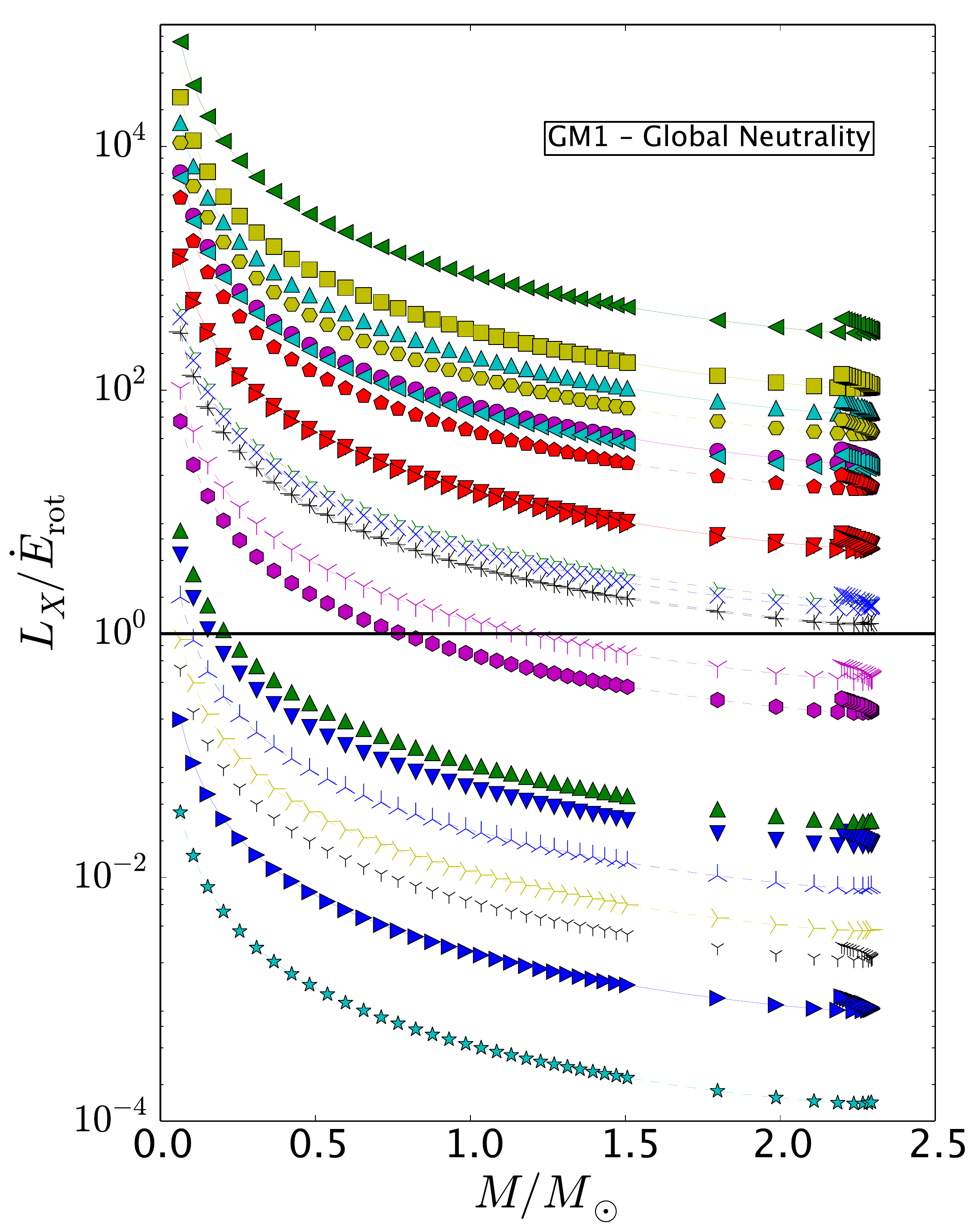}\includegraphics[width=0.5\hsize,clip]{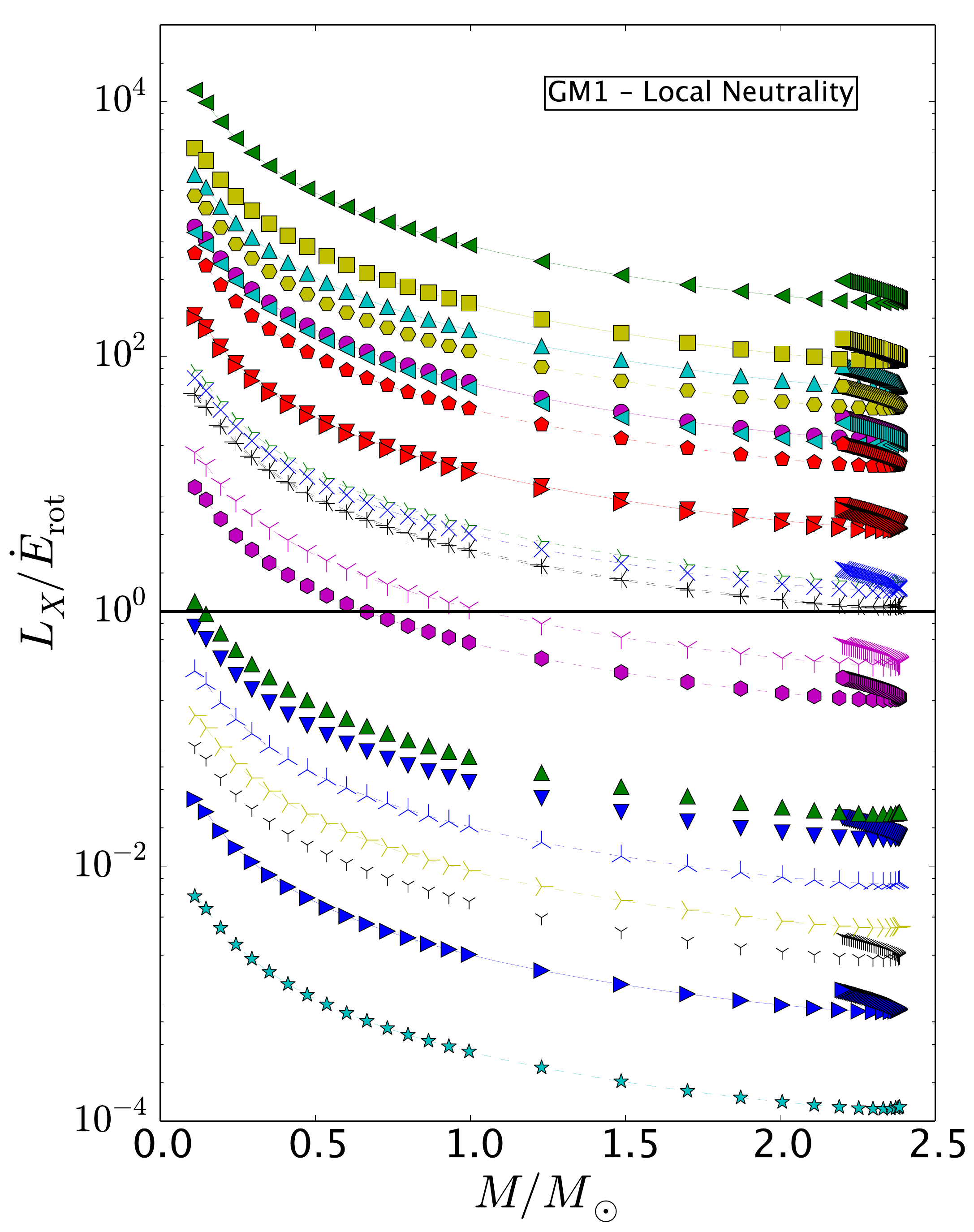}
\caption{(Color online) Radiation efficiency $L_X/\dot{E}_{\rm rot}$ as a function of the NS mass (in solar masses), for the global (left panel) and local (right panel) neutrality cases.}\label{fig:LxEdot}
\end{figure*}

Figure~\ref{fig:LxEdot} shows the X-ray luminosity to rotational energy loss ratio as a 
function of the NS mass, for both global and local charge neutrality configurations. We can see from these figures that nine out of the twenty three SGR/AXPs could have masses in which $L_X<\dot{E}_{\rm rot}$, and therefore they could be explained as ordinary rotation-powered NSs. Such sources are: Swift J1834.9--0846, PSR J1846--0258, 1E 1547.0--5408, SGR J1745-2900, XTE J1810--197, PSR J1622--4950, SGR 1627--41, SGR 0501+4516, CXOU J171405.7381031~(see Table~\ref{tab9sources}). In view of the proximity of some of the sources to the line $L_X/\dot{E}_{\rm rot} = 1$ (e.g., SGR 1900+14, SGR 0418+5729, and Swift J1822.3--1606), and the currently poorly constrained determination of the distance to the sources, there is still the possibility of having additional sources as rotation powered NSs.

In this line, two sources are particularly interesting, namely SGR 1900+14 and SGR 1806-20, which appear very close (but above) to the limit of becoming rotation-powered NSs. The soft X-rays spectra of SGRs and AXPs are usually well fitted by a blackbody + power-law spectral model \citep[see e.g.,][]{2008A&ARv..15..225M}. The blackbody temperature is usually of the order $k_B T\sim 0.5$~keV and surface radii of the emitting region are $\sim 1$~km. In the case of a NS, one could interpret such a thermal component as due to the surface  temperature of the NS, namely associated with the thermal reservoir of the star. The power-law component has instead a non-thermal nature and must be due to magnetospheric processes which are connected with the rotational energy reservoir in a rotation-powered star. Within this interpretation, the request that the rotational energy loss pays also for the contribution of the thermal component of the luminosity is unnecessarily rigorous. Thus, the above ratio $L_X/\dot{E}_{\rm rot}$ becomes an upper limit to the actual efficiency for the conversion of rotational into electromagnetic energy. We can now apply this interpretation, for the sake of example, to the above two sources.

\emph{SGR 1900+14:} the blackbody component of the spectrum is characterized by $k_B T_{\rm BB} = 0.47$~keV, and a surface radius of $R_{\rm BB} = 4.0$~km, assuming a distance of 15~kpc \citep{2006ApJ...653.1423M}. The total (blackbody + power-law) flux in the 2--10~keV energy band is $F_X=4.8\times 10^{-12}$~erg~cm$^{-2}$~s$^{-1}$. With the above data, we infer that the blackbody and the power-law components contribute respectively 28\% and 72\% to the total flux. Namely, we have $F^{\rm BB}_X=0.28 F_X$ and $F^{\rm PL}_X=0.72 F_X$. This leads to $L^{\rm PL}_X = 9.3\times 10^{34}$~erg~s$^{-1}$.

\emph{SGR 1806--20:} in this case we have $k_B T_{\rm BB} = 0.55$~keV and $R_{\rm BB} = 3.7$~km, assuming a distance of 15~kpc \citep{2007A&A...476..321E}. For this source $F_X=1.8\times 10^{-11}$~erg~cm$^{-2}$~s$^{-1}$, and we infer $F^{\rm BB}_X=0.16 F_X$ and $F^{\rm PL}_X=0.84 F_X$. This leads to $L^{\rm PL}_X = 4.1\times 10^{35}$~erg~s$^{-1}$. If we use instead the revised distance of 8.7~kpc \citep{2008MNRAS.386L..23B}, we have $L^{\rm PL}_X = 1.4\times 10^{35}$~erg~s$^{-1}$.

\begin{figure*}
\centering
\includegraphics[width=0.5\hsize,clip]{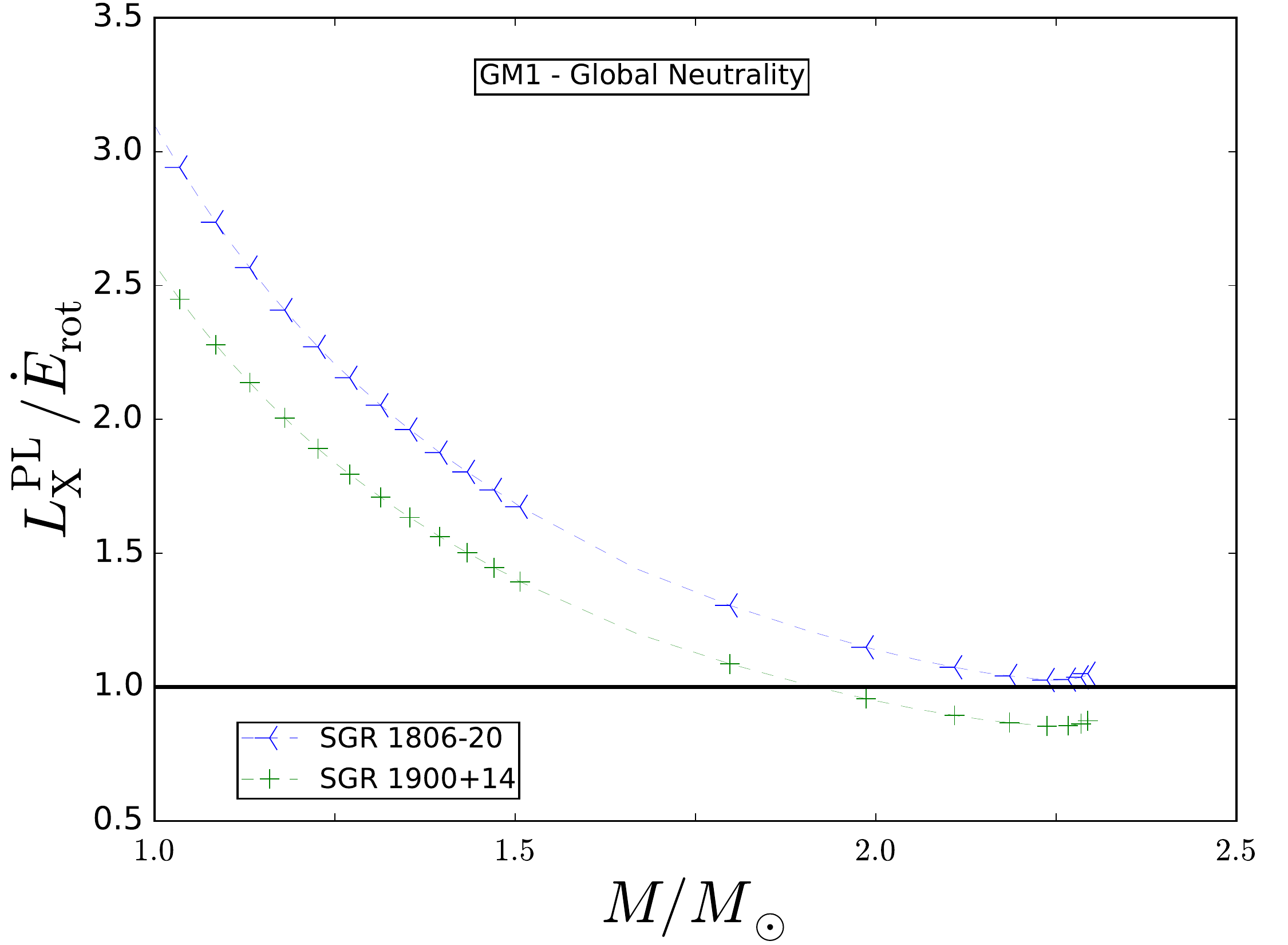}\includegraphics[width=0.5\hsize,clip]{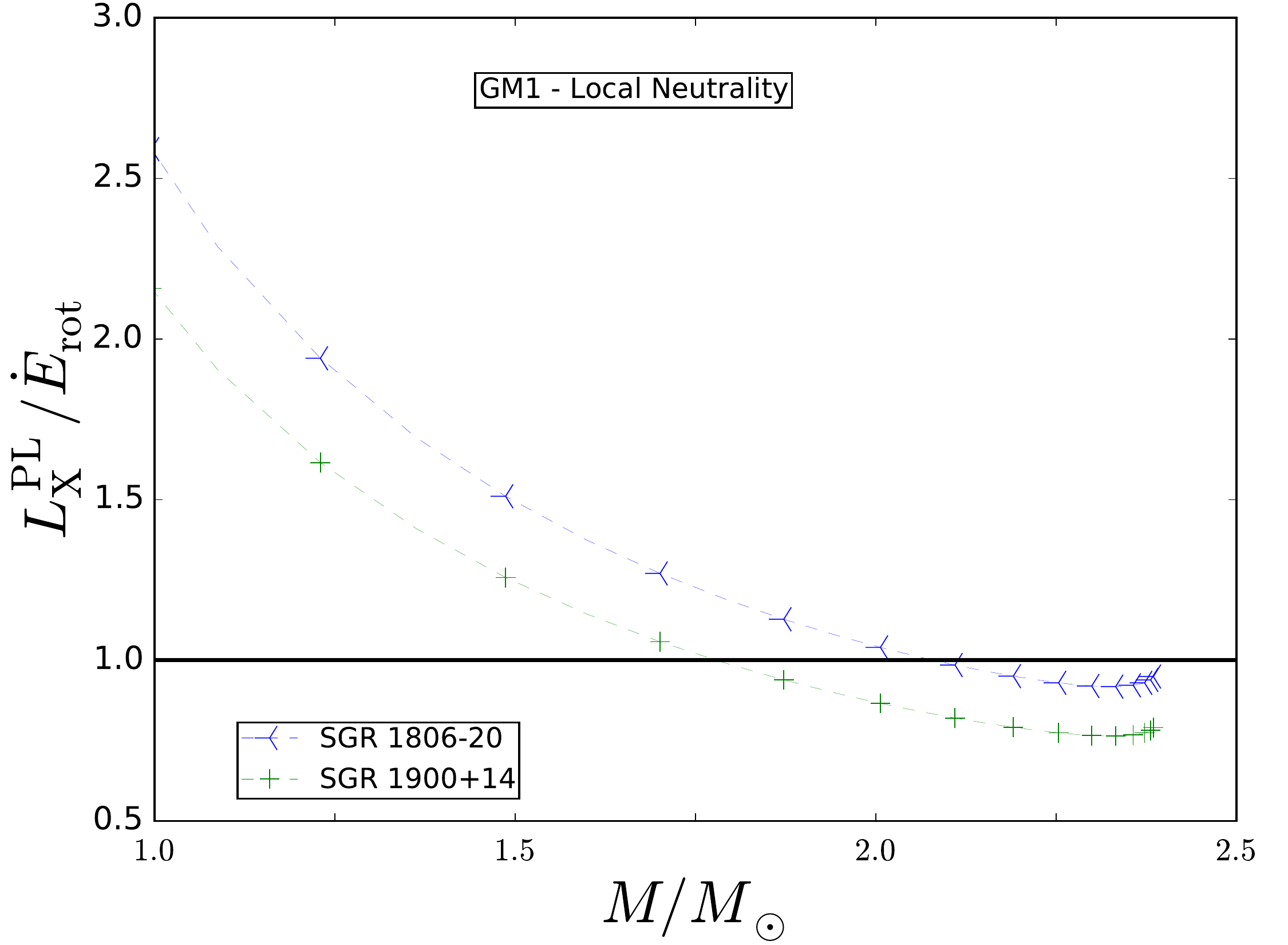}
\caption{(Color online) Radiation efficiency $L_X/\dot{E}_{\rm rot}$ as a function of the NS mass (in solar masses), for the global (left panel) and local (right panel) neutrality cases.}\label{fig:LxEdotPL}
\end{figure*}

Figure~\ref{fig:LxEdotPL} shows the ratio $L^{\rm PL}_X/\dot{E}_{\rm rot}$ as a function of the NS mass in the case of SGR 1900+14 and SGR 1806--20, adopting the GM1 EOS and assuming a distance of 15~kpc for both sources. It is clear from this analysis the importance of identifying the different contributions to the emission of the object. There is no doubt that the subtraction of the contribution from the thermal reservoir to the total flux in soft X-rays can be important for the correct identification of the nature of these sources: now there is a range of masses for which the luminosity to rotational energy loss ratio becomes lower than unity. Again, it is worth to recall that there are still additional effects which could improve the above analysis: (i) the distance to the sources are not known accurately; (ii) the spectrum could be equally well-fitted by a different spectral model such as a double blackbody which would have a different interpretation; (iii) the NS EOS is still unknown and so the moment of inertia and radius for a given mass might be different. These effects might lead to a different value of the luminosity, and of the contributions of thermal and rotational energy reservoirs to it. Clearly, the above analysis can be extended to all the other SGRs and AXPs, and in the case of the nine sources already identified with $L_X/\dot{E}_{\rm rot} < 1$, it will further diminish their radiation efficiency.

It is now appropriate to discuss the non-thermal hard X-ray emission (above 10~keV) in SGRs/AXPs which has been observed by some missions like {\it RXTE, INTEGRAL, Suzaku} and {\it NuSTAR}. 
First we discuss the observations of the above two sources which could in principle join the possibly rotation-powered group. Adopting a distance of 15~kpc, SGR 1900+14 has an observed 20--100~keV band luminosity of $L^{\rm hard}_X = 4\times 10^{35}$~erg~s$^{-1}$ \citep{2006A&A...449L..31G}. This implies $L^{\rm hard}_X\approx 4.3 L^{\rm PL}_X$, so a total X-ray luminosity (hard $+$ soft) $L_x = 5.3 L^{\rm PL}_X \approx 4.9\times 10^{35}$~erg~s$^{-1}$. SGR 1806--20 has a 20--100~keV band flux three times higher than the one of SGR 1900+14 \citep{2006A&A...449L..31G}, thus assuming also a distance of 15~kpc for this source we obtain $L^{\rm hard}_X\approx 2.9 L^{\rm PL}_X$, so a total X-ray luminosity (hard $+$ soft) $L_x = 3.9 L^{\rm PL}_X \approx 1.6\times 10^{36}$~erg~s$^{-1}$. This means that the points in Figure~\ref{fig:LxEdotPL} would shift 5 and 4 times up respectively and therefore there will be no solution for these sources as rotation-powered, unless their distances are poorly constrained. In this line it is worth mentioning that the distance to these sources has been established via their possible association with star clusters \citep[see][for details]{2000ApJ...533L..17V,2004A&A...419..191C}.

\begin{figure*}
\centering
\includegraphics[width=0.5\hsize,clip]{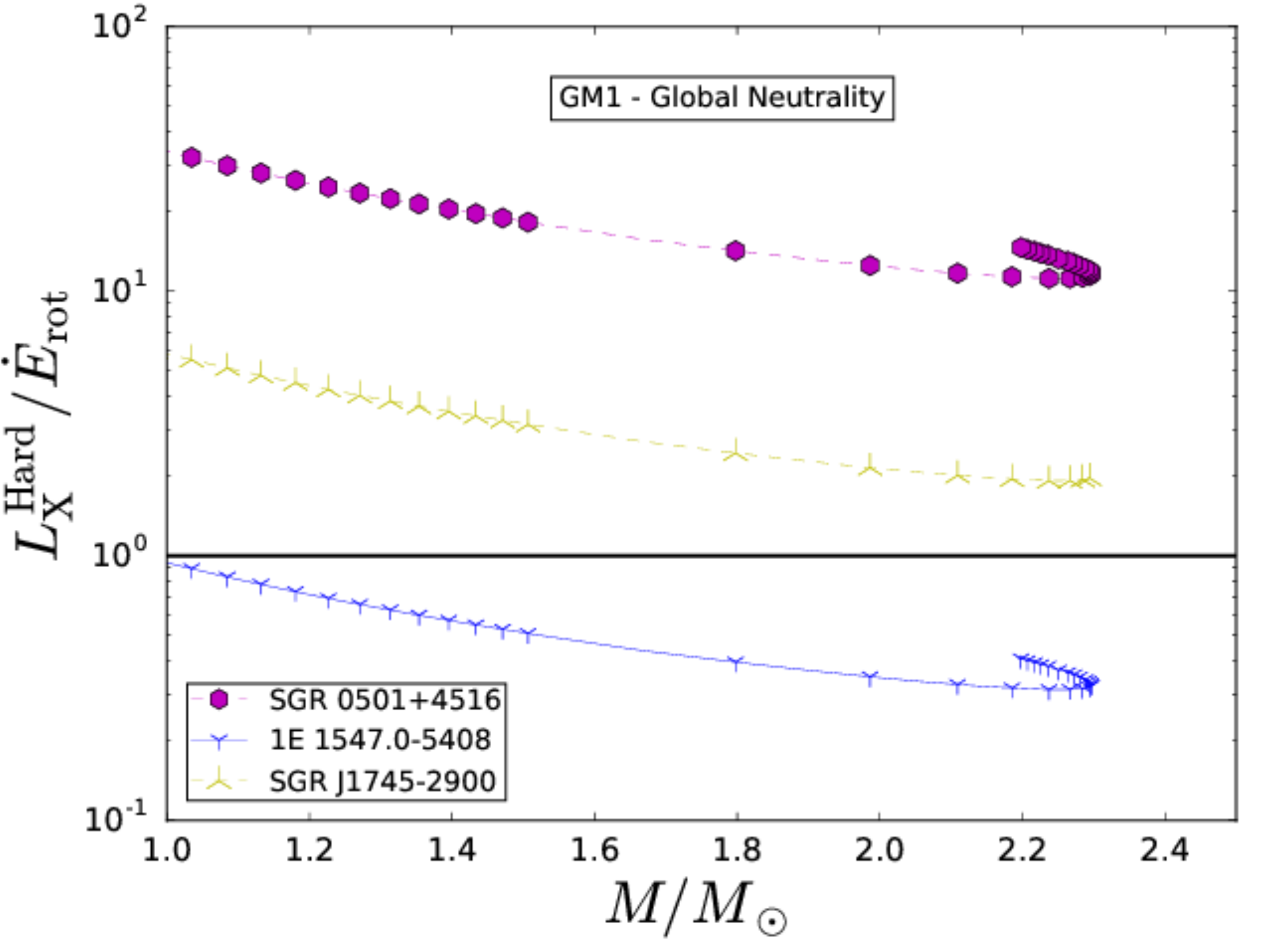}\includegraphics[width=0.5\hsize,clip]{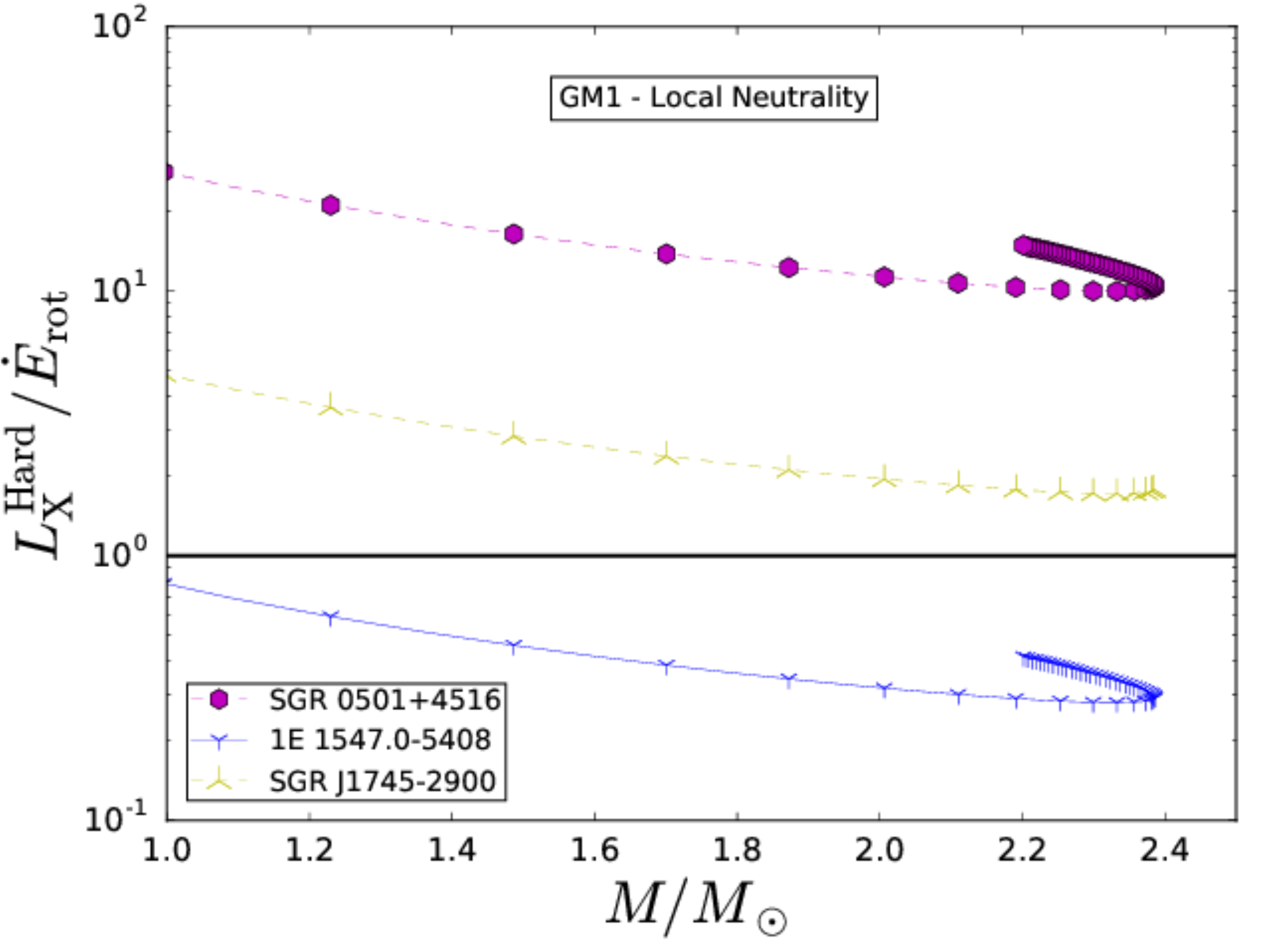}
\caption{(Color online) Radiation efficiency $L_X^{\rm Hard}/\dot{E}_{\rm rot}$ as a function of the NS mass (in solar masses), for the global (left panel) and local (right panel) neutrality cases.}\label{fig:LxEdotHard}
\end{figure*}

From the set of nine potential rotation-powered sources, only three ones have persistent hard X-ray emission (see Table~\ref{tab9sources}): SGR 0501+4516, 1E 1547.0--5408 and SGR J1745--2900. For these sources we can see the hard X-ray luminosity in the 20--150~keV band overcomes the soft X-ray contribution to the luminosity respectively by a factor 50, 149 and 527. Figure~\ref{fig:LxEdotHard} shows the ratio $L^{\rm Hard}_X/\dot{E}_{\rm rot}$ as a function of the NS mass in the case of SGR 0501+4516, 1E 1547.0--5408 and SGR J1745--2900. We can see that, after including the hard X-ray component in these three sources, 1 E1547.0--5408 stands still below the line $L_X/\dot{E}_{\rm rot} = 1$, while the other two sources appear above it.

The existence of persistent hard X-ray emission provides new constraints on the emission models for SGRs/AXPs since, as in ordinary pulsars, the higher the energy band the higher the luminosity, namely their luminosities can be dominated by hard X-rays and/or gamma-rays. At the present, the mechanisms responsible for the hard energy emission is still poorly understood, what causes the hard X-ray tails is still an open issue. In this respect it is worth mentioning that, since these sources are also associated with supernova remnants (see Table~\ref{tab9sources} and Sec.~\ref{sec:7}), the emission in hard X and/or gamma-rays could be contaminated by the remnant emission. The disentanglement of the contributions of the remnant and the central pulsar to the total emission is an interesting issue to be explored, in addition to the confirmation of the estimated distances. If the above numbers will be confirmed, then the number of rotation-powered SGRs/AXPs becomes seven.

\begin{table*}
\centering
{\scriptsize
\caption{{Some properties of nine SGRs/AXPs potential rotation-powered NSs. Column 1: source name. Column 2: rotation period $P$ in units of seconds. Column 3: spin-down rate $\dot{P}$ in units of $10^{-11}$. Column 4: source distance in units of kpc. Column 5: X-ray luminosity in the 2--10~keV band in units of $10^{33}$~erg~s$^{-1}$. Column 6: hard X-ray luminosity in the 20--150~keV band in units of $10^{33}$~erg~s$^{-1}$. Column 7: radio luminosity per solid angle at the frequency $f_0 = 1.4$~GHz, i.e. $L_{\rm radio} = S_{1.4} d^2$ in units of $10^{28}$~sr$^{-1}$~erg~s$^{-1}$ where $S_{1.4}$ is the measured flux density at $f_0$. In the case of SGR J1745-2900 we report the luminosity per beam at the frequency 41~GHz according to \citet{2015ApJ...811L..35Y}. In the columns 8--11 we report respectively if the source has reported association with SNR, observed glitches, outbursts, and if it is considered as a transient X-ray lumminosity (in the sense explained in Sec.~\ref{sec:7}). Data have been taken from the McGill catalog (\citealp{2014ApJS..212....6O}; see http://www.physics.mcgill.ca/~pulsar/magnetar/main.html).}}\label{tab9sources}
\begin{tabular}{l|c|c|c|c|c|c|c|c|c|c}
\hline
Source &$P$ &$\dot{P}$ & $d$ & $L_X$ & $L^{\rm hard}_X$& $L_{\rm radio}$ & SNR Assoc.&Obs. Glitches & Burst & Transient\\ & (s) & & (kpc) & ($10^{33}$~erg~s$^{-1}$) & ($10^{33}$~erg~s$^{-1}$) & ($10^{28}$~sr$^{-1}$~erg~s$^{-1}$) & & & &
\\
						\hline
						SGR 0501+4516 &5.8 &0.59 & 2 &0.81 & 40.2 & -- & HB 9 (?) &No &Yes&Yes\\
						1E 1547.0--5408  &2.07 &4.77 & 4.5 & 1.3 & 193.9 & 1.19 & G327.24-0.13 &Yes &Yes&Yes\\
						PSR J1622-4950  &4.33 &1.7 & 9 & 0.44 & -- & 5.18 & G333.9+0.0 &No &No&Yes\\
						SGR 1627-41  &2.59 &1.9 & 11 & 3.6 & -- & -- & G337.0-0.1 &No &Yes&Yes\\
						CXOU J171405.7-381031  &3.8 &6.4& 13.2 &56 & -- & -- & CTB 37B &No &No&No\\
						SGR J1745-2900  &3.76 &1.38 & 8.5 &0.11 & 57.9 & 84.6 & --	 &No &Yes&Yes\\
						XTE J1810-197  &5.54 &0.77 & 3.5 & 0.043 & -- & 0.98 & -- &No &Yes&Yes\\
						Swift J1834.9-0846  &2.48 &0.79 & 4.2 &0.0084 & -- & -- & W41 &No &Yes&Yes\\
						PSR J1846-0258  &0.33 &0.71 & 6 & 19 & -- & -- & Kes 75 &Yes &Yes&No\\
                        \hline
					\end{tabular}
}
\end{table*}

\section{Glitches and bursts in SGRs/AXPs}\label{sec:5}

We have shown in the last section that nine (and up to eleven) of the twenty three SGR/AXPs are potential rotation-powered NSs. Once the possible rotation-power nature of the source is established, one expects that also the transient phenomena observed in these sources could be powered by rotation. Based on that idea, we here discuss a possible glitch-outburst connection. Thus, it is interesting to scrutinize the outburst data of SGR/AXPs, to seek for associated glitches, and check if the energetics of its bursting activity could be explained by the gain of rotational energy during an associated (observed or unobserved) glitch.

In a glitch, the release of the accumulated stress leads to a sudden decrease of the moment of inertia and, via angular momentum conservation,
\begin{equation}
J=I\Omega= (I + \Delta I)(\Omega+\Delta\Omega)={\rm constant},
\end{equation}
to a decrease of both the rotational period (spin-up) and the radius, i.e.
\begin{equation}
\frac{\Delta I}{I}=2\frac{\Delta R}{R}=\frac{\Delta P}{P}= -\frac{\Delta\Omega}{\Omega}.
\end{equation}
The sudden spin-up leads to a gain of rotational energy
\begin{equation}\label{eq:DeltaErot}
\Delta E_{\rm rot} = - \frac{2 \pi^2 I}{P^2}\frac{\Delta P}{P},
\end{equation}
which is paid by the gravitational energy gain by the star's contraction \citep{2012PASJ...64...56M}.

It is important to start our analysis by recalling the case of PSR J1846--0258, which has $L_X<\dot{E}_{\rm rot}$ even when fiducial NS parameters are adopted. The importance of this source relies on the fact that, although it is recognized as rotation-powered NS, it has been classified as SGR/AXP \citep{2014ApJS..212....6O} owing to its outburst event in June 2006 \citep{2008Sci...319.1802G}. In view of the possible NS rotation-power nature of PSR, \citet{2012PASJ...64...56M} explored the possibility that the outburst energetics $(3.8$--$4.8)\times 10^{41}$~erg \citep{2008ApJ...678L..43K} can be explained by the rotational energy gained by a NS glitch. It was there found that a glitch with fractional change of period $|\Delta P|/P \sim (1.73$--$2.2)\times 10^{-6}$ could explain the outburst of 2006. This theoretical result is in full agreement with the observational analysis by \citet{2009A&A...501.1031K}, who showed that indeed a major glitch with $|\Delta P|/P\sim (2$--$4.4)\times10^{-6}$ is associated with the outburst.
Very recently,~\citet{2016arXiv160801007A} reported another example of an X-ray outburst from a radio pulsar, PSR J1119--6127, which also has $L_X<\dot{E}_{\rm rot}$. This source is similar to the rotation-powered pulsar PSR J1846--0258. The pulsar's spin period $P=0.407$~s and spin-down rate $\dot{P}=4.0\times 10^{-12}$ imply a dipolar surface magnetic field $B=4.1\times10^{13}$~G adopting fiducial values. It is clear from Figs.~\ref{figure1} and \ref{figure2} that also in this case the magnetic field would become undercritical for realistic NS parameters.

We follow this reasoning and proceed to theoretically infer the fractional change of rotation period, $|\Delta P|/P$, which explains the energetics of the bursts of the family of SGR/AXPs with $L_X<\dot{E}_{\rm rot}$ presented in this work. We do this by assuming that $|\Delta E_{\rm rot}|$, given by Eq.~(\ref{eq:DeltaErot}), equals the observed energy of the burst event, $E_{\rm burst}$, namely 
\begin{equation}\label{eq:DeltaP}
\frac{|\Delta P|}{P} = \frac{E_{\rm burst} P^2}{2\pi^2 I}.
\end{equation}

\begin{figure*}
\centering
\includegraphics[width=0.5\hsize,clip]{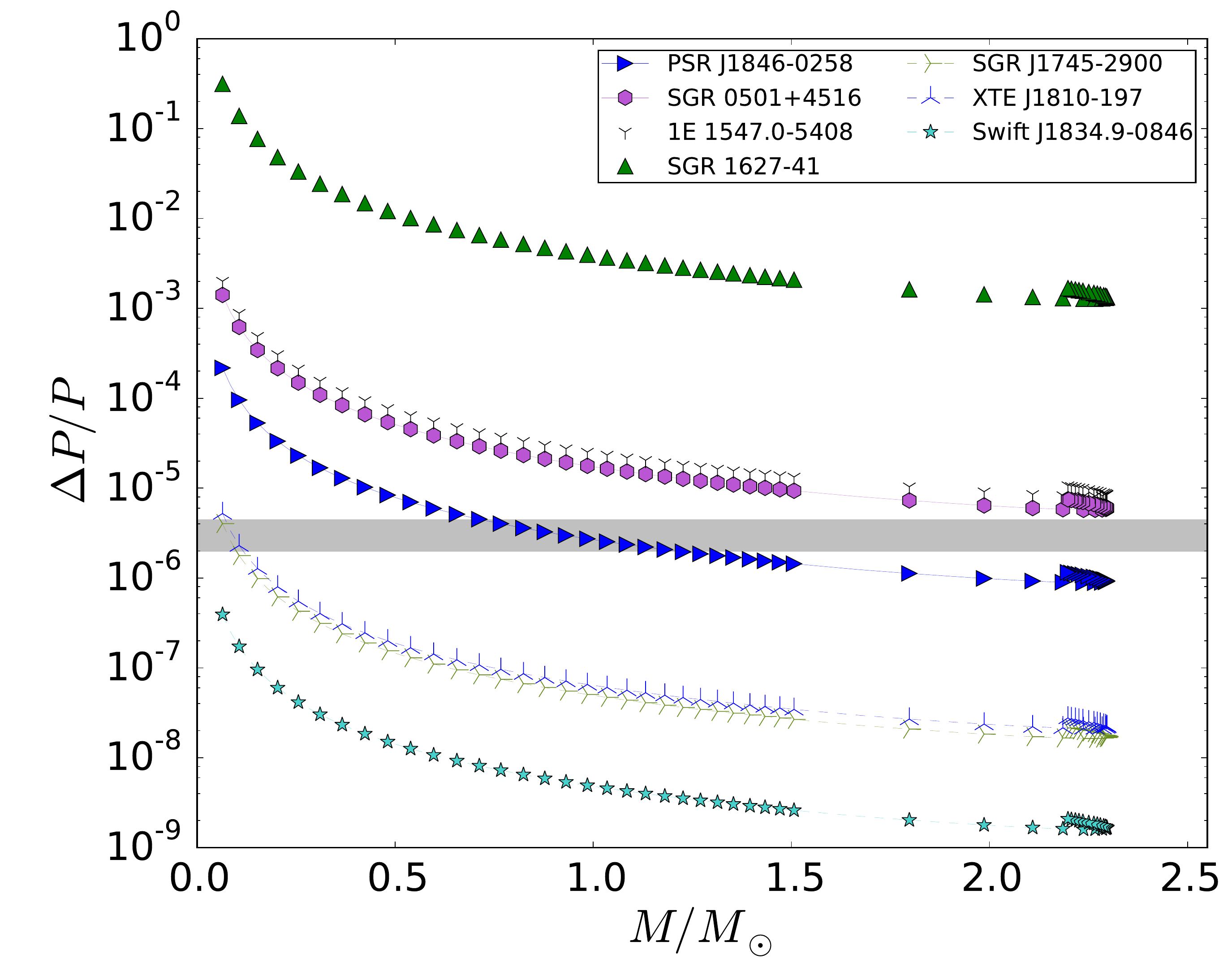}\includegraphics[width=0.5\hsize,clip]{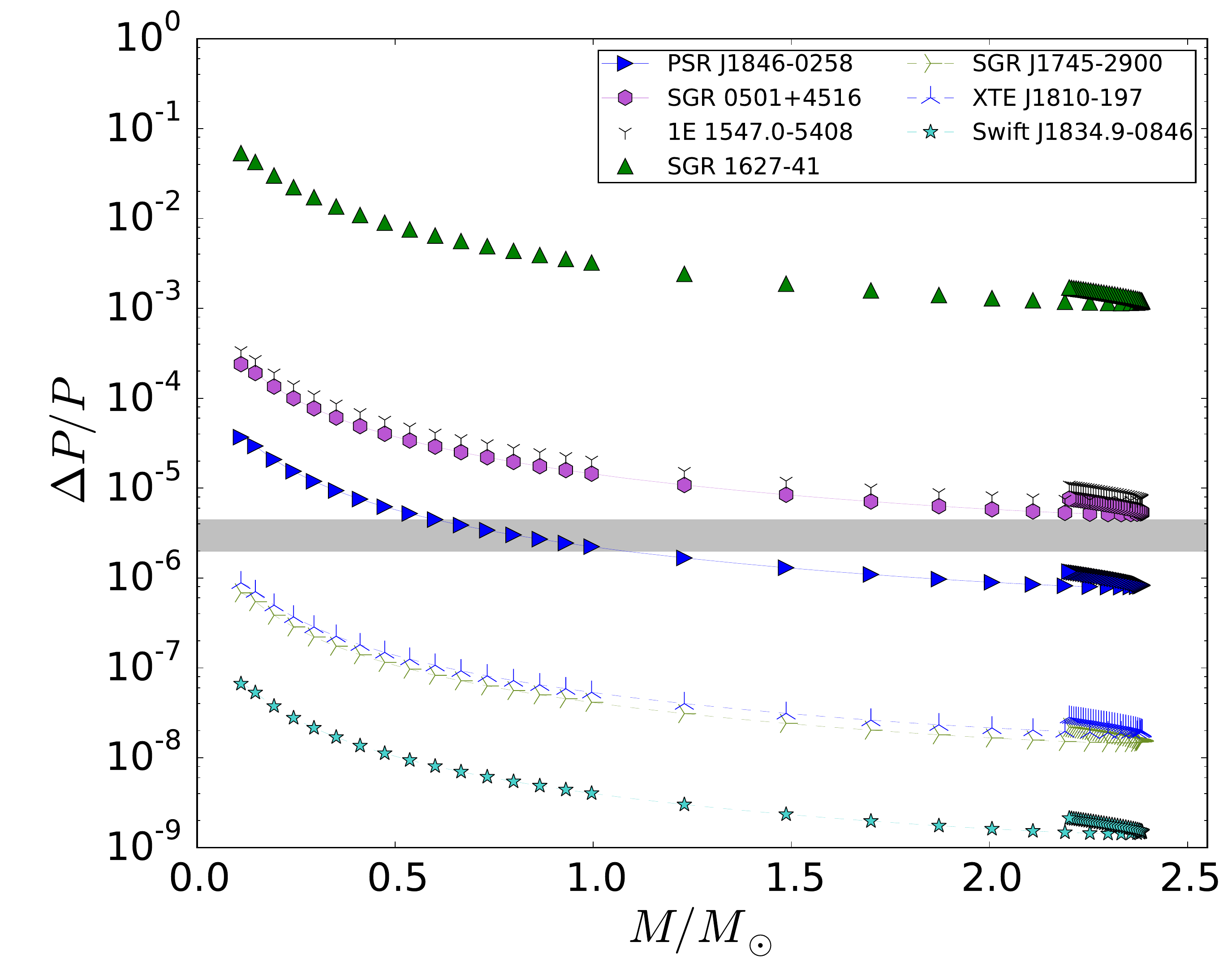}
\caption{(Color online) Inferred fractional change of rotation period during the glitch, $\Delta P/P$, obtained by equating
the rotational energy gained during the glitch, $\Delta E_{\rm rot}$, to the energy of the burst, for globally neutral
(left panel) and locally neutral (right panel) NSs. In this example the NS obeys the GM1 EOS. The gray-shaded area 
corresponds to the value of $|\Delta P|/P$ in the observed glitch of PSR J1846--0258 in June 2006 \citep{2009A&A...501.1031K}.}\label{fig:glitch}
\end{figure*}

From the set of nine  potential rotation-powered sources, only two ones have glitches detected: PSR J1846--0258 which has been discussed above and 1E 1547.0--5408 with $|\Delta P|/P\approx 1.9\times 10^{-6}$~\citep{2012ApJ...748..133K}. In particular, figure~\ref{fig:glitch} shows the value of $|\Delta P|/P$  obtained from equation~(\ref{eq:DeltaP}) as a function of the NS mass for PSR J1846--0258 (a similar analysis can be applied to the radio pulsar PSR J1119--6127, whose timing analysis presented in \citealp{2016arXiv160801007A} suggests the pulsar had a similar-sized spin-up glitch with $|\Delta P|/P \sim 5.8\times 10^{-6}$).
Indeed a minimum mass for the NS can be established for the sources by requesting that: 1) the entire moment of inertia is involved in the glitch and 2)  the theoretical value of $|\Delta P|/P$
coincides with the observed value. We obtain a minimum mass for PSR J1846--0258, $M_{\rm min} = 0.72~M_\odot$ and $M_{\rm min} = 0.61~M_\odot$, for the global and local charge 
neutrality cases, respectively. On the other hand, if we substitute the moment of inertia $I$ in equation~(\ref{eq:DeltaP}) by $I_{\rm glitch}=\eta I$ where $\eta \leq 1$, being $I_{\rm glitch}$ the moment of inertia powering the glitch, then we can obtain a lower limit for the parameter $\eta$: we obtain $\eta=0.20$ and $\eta=0.18$ for the global and local charge neutrality cases, respectively. Tables~\ref{tab:table2} and \ref{tab:table3} show the theoretically predicted value of $|\Delta P|/P$ for the seven sources with known bursts energy, assuming the mass of the NS is larger than $1~M_\odot$ and $\eta=1$, in the cases of global and local charge neutrality, respectively.

\begin{table*}
\centering
\caption{Predicted values of $|\Delta P|/P$ assuming rotation-powered NSs - Global charge neutrality case}\label{tab:table2}
\begin{tabular}{l|c|c|c}
\hline
Source name & Year of burst & Total isotropic burst energy~(erg) & Predicted $|\Delta P|/P$ for $M>1\,M_\odot$\\
\hline
PSR J1846--0258& $2006$  & $4.8\times 10^{41}$     & $8.8\times 10^{-7} - 2.6\times 10^{-6}$      \\
1E 1547.0--5408  & $2009$  & $1.1\times 10^{41}$     & $8.1\times 10^{-6} - 2.4\times 10^{-5}$ \\
XTE J1810--197 & $2004$  & $4.0\times 10^{37}$     & $2.1\times 10^{-8} - 6.3\times 10^{-8}$ \\
SGR 1627--41 & $1998$  & $1.0\times 10^{41}$     &$1.0\times 10^{-5} - 3.8\times 10^{-5}$  \\
SGR 0501+4516 & $2008$  & $1.0\times 10^{40}$     & $5.7\times 10^{-6} - 1.7\times 10^{-5}$ \\
Swift J1834.9--0846 & $2011$  & $1.5\times 10^{37}$     &$1.6\times 10^{-9} - 4.8\times 10^{-9}$ \\
SGR 1745--2900 & $2013$  & $6.7\times 10^{37}$     & $1.61\times 10^{-8} - 4.9\times 10^{-8}$\\
\hline
\end{tabular}
\end{table*}

\begin{table*}
\centering
\caption{Predicted values of $|\Delta P|/P$ assuming rotation-powered NSs - Local charge neutrality case}\label{tab:table3}
\begin{tabular}{l|c|c|c}
\hline
Source name & Year of burst  & Total isotropic burst energy~(erg) & Predicted $|\Delta P|/P$ for $M>1\,M_\odot$\\
\hline
PSR J1846--0258& $2006$  & $4.8\times 10^{41}$     & $7.9\times 10^{-7} - 2.2\times 10^{-6}$      \\
1E 1547.0--5408  & $2009$  & $1.1\times 10^{41}$     & $7.2\times 10^{-6} - 2.0\times 10^{-5}$ \\
XTE J1810--197 & $2004$  & $4.0\times 10^{37}$     & $1.9\times 10^{-8} - 5.3\times 10^{-8}$ \\
SGR 1627--41 & $1998$  & $1.0\times 10^{41}$     &$1.1\times 10^{-5} - 3.2\times 10^{-5}$  \\
SGR 0501+4516 & $2008$  & $1.0\times 10^{40}$     & $5.0\times 10^{-6} - 1.4\times 10^{-5}$ \\
Swift J1834.9--0846 & $2011$  & $1.5\times 10^{37}$     &$1.4\times 10^{-9} - 3.9\times 10^{-9}$ \\
SGR 1745--2900 & $2013$  & $6.7\times 10^{37}$     & $1.4\times 10^{-8} - 4.1\times 10^{-8}$\\
\hline
\end{tabular}
\end{table*}

Table~\ref{tab:table3} shows that from the nine potentially rotation-powered sources, two have a firmly established glitch-outburst connection. For the other sources there are two possibilities. 1) The glitch could be missed because absence of timing monitoring of the source prior to the burst, as it is certainly the case of the SGRs/AXPs discovered from an outburst. 2) The source timing was monitored and indeed there is no glitch associated with the outburst. In this case, it remains open the possibility that the outburst could be of magnetospheric origin. 3) There are also observed glitches without associated outburst activity \citep[see, e.g.,][]{2012ApJ...750L...6P}. It is worth mentioning that a recent systematic analysis of the glitch-outburst connection in five AXPs by \citet{2014ApJ...784...37D} concluded (amongst other important results): 1) glitches associated and not associated with outbursts or radiative changes show similar timing properties, namely outburst activity is not necessarily associated with large glitches; and 2) all glitches observed point to have their origin in the stellar interior. The second conclusion gives observational support to our theoretical interpretation of  glitches as a phenomenon associated to cracking occurring in the NS interior. Whether a glitch can or not lead to observable radiative changes depends on specific properties of the phenomenon such as the energy budget and the localization of the event in the star's interior \citep{2014ApJ...784...37D}, as well as on the efficiency in converting mechanical energy into radiation. The first two features have been here analyzed through $\Delta E_{\rm rot}$ and the parameter $\eta$, the latter which defines $I_{\rm glitch}$, the amount of moment of inertia involved in the glitch. 

Thus, the glitch-outburst connection remains one of the most interesting problems of SGR/AXP physics and astrophysics. There are still several issues which need to be addressed both from systematic observational analyses and from theoretical point of view of NS physics.

\section{Possible additional evidence}\label{sec:7}

We have shown above for the nine potential rotation-powered SGRs/AXPs that, when timing observations allowed for the glitch/outburst connection identification, the rotational energy gain in the glitch can explain the outburst energetics. This characteristic is expected from a rotation-powered object. 

We discuss now three additional pieces of astrophysical evidence pointing to a rotation-power nature of these sources. First, we note that four of the above nine sources, namely 1E 1547.0--5408, SGR J1745--2900, XTE J1810--197, and PSR J1622--4950, are the only SGR/AXPs with detected radio emission \citep[see, e.g.,][]{2005ApJ...632L..29H,2006Natur.442..892C,2006HEAD....9.0603H,2007MNRAS.377..107K,2007ApJ...663..497C,2007ApJ...666L..93C,2008ApJ...679..681C,2010ApJ...721L..33L,2012MNRAS.422.2489L,2013Natur.501..391E,2014ApJS..212....6O,2015JPhCS.630a2015L,2015ApJ...811L..35Y}. 
This property, expected in ordinary rotation-powered pulsars, is generally absent in SGR/AXPs. As discussed in \citet{2007MNRAS.377..107K}, the radio emission of SGRs/AXPs and normal radio pulsars shows differences but also similarities, e.g. the case of XTE J1810-197. We show in Table~\ref{tab9sources} the observed radio luminosity per solid angle at the 1.4~GHz frequency. For all of them we have $L_{\rm radio} \ll L_X$, a feature also observed in ordinary pulsars. The continuous observation, as well as theoretical analysis and comparison of the radio emission of rotating radio transients (RRATS), high-B pulsars, SGRs/AXPs, and ordinary radio pulsars will allow us to understand the NS properties leading to the differences and similarities of the radio emission of these sources. New observational capabilities such as the ones of the Square Kilometer Array (SKA) expect to give also important contributions in this direction \citep{2015aska.confE..39T}.

In order to understand better the nature of these nine SGRs/AXPs it is worth to seek for additional emission features which could distinguish them from the rest of the sources. In this line we would like to point out that, at present, eleven SGRs/AXPs have been identified as \emph{transient} sources, i.e.~sources which show flux variations by a factor $\sim$(10-1000) over the quiescent level \citep[see, e.g.,][]{2015RPPh...78k6901T}, in timescales from days to months. Such a variations are usually accompanied by an enhancement of the bursting activity. Seven of the nine potentially rotation-powered sources shown in Table~\ref{tab9sources} are transient sources. Therefore it is possible to identify common features in these sources, although more observational and theoretical investigation is needed. For instance, the above shows that all radio SGR/AXPs are rotation-powered sources and have a transient nature in the X-ray flux. The theoretical analysis of the evolution of the X-ray flux can constrain the properties of the NS and the emission geometry, for instance the angles between the rotation axis, the line of sight and the magnetic field \citep[see, e.g.,][]{2010ApJ...722..788A}. Such constraints can help in constraining, at the same time, the properties of the radio emission. Such an analysis is however out of the scope of the present work and opens a window of new research which we plan to present elsewhere. 

There is an additional observational property which support to a NS nature for the nine sources of Table~\ref{tab9sources}, namely six of them have possible associations with supernova remnants (SNRs). As it is summarized in Table~\ref{tab9sources}, Swift J1834.9--0846 has been associated with SNR W41 (see, \citealp{2016arXiv160406472Y}, for the detection of a wind nebula around this source); PSR J1846--0258 with SNR Kes75; 1E 1547.0--5408 with SNR G327.24--0.13; PSR J1622--4950 with SNR G333.9+0.0; SGR 1627--41 with NR G337.0-0; and CXOU J171405.7--381031 with SNR CTB37B \citep{2014ApJS..212....6O}. If these associations will be fully confirmed, then it is clear that the NS was born from the core-collapse of a massive star which triggered the SN explosion. Further analysis of the supernova remnant and/or pulsar wind nebulae energetics and emission properties is needed to check their consistency with the rotation-powered nature of the object at the center.

\section{Conclusions}\label{sec:8}

We considered in this work the possibility that some SGRs and AXPs can be rotation-powered NSs exploring the allowed range of realistic NS structure parameters for the observed rotation periods of SGRs/AXPs, instead of using only fiducial parameters $M=1.4M_\odot$, $R=10$~km, and $I=10^{45}$~g~$\rm cm^2$. We obtained the NS properties from the numerical integration of the general relativistic axisymmetric equations of equilibrium for EOS based on relativistic nuclear mean-field models both in the case of local and global charge neutrality. We thus calculated the rotational energy loss, $\dot{E}_{\rm rot}$, (hence to a radiation efficiency $L_X/\dot{E}_{\rm rot}$) as a function of the NS mass. In addition, we estimate the surface magnetic field from a general relativistic model of a rotating magnetic dipole in vacuum.

Based on the above, the following conclusions can be drawn:
\begin{itemize}
\item 
Fiducial parameters overestimate both the radiation efficiency and the surface magnetic field of pulsars.
\item
The X-ray luminosity of nine sources shown in Table~\ref{tab9sources}, i.e. Swift J1834.9--0846, PSR J1846--0258, 1E 1547.0--5408, SGR J1745--2900, XTE J1810--197, PSR J1622--4950, SGR 1627--41, SGR 0501+4516, CXOU J171405.7381031, can be explained via the loss of rotational energy of NSs (see Fig.~\ref{fig:LxEdot}). Thus, they fit into the family of ordinary rotation-powered pulsars. 
\item
For the above nine sources, we obtained lower mass limits from the request $\dot{E}_{\rm rot}\geq L_X$. 
\item 
We show that, if the thermal reservoir of the NS is the responsible of the blackbody component observed in soft X-rays, both SGR 1900+14 and SGR 1806-20 join the above family of rotation-powered NSs since the rotational energy loss is enough to cover their non-thermal X-ray luminosity.  This implies that up to 11 SGR/AXPs could be rotation-powered pulsars. This argument could be in principle also applied to the other sources, lowering further their radiation efficiency $L_X/\dot{E}_{\rm rot}$.
\item 
Thus we argue that the observational uncertainties in the determination of the distances and/or luminosities, as well as the uncertainties in the NS nuclear EOS, as well as the different interpretations of the observed spectrum leave still room for a possible explanation in terms of spin-down power for additional sources. It is worth mentioning that the distance for some sources has been established via their association with supernova remnants (SNRs), as pointed out in Table~\ref{tab9sources}.
\item We then proceeded to discuss the observed emission in hard X-rays (in the 20--150~keV band) first in both SGR 1900+14 and SGR 1806-20. Including this contribution the luminosity increases up to a factor 5 and 4 respectively for each source, leaving no room for them as rotation-powered sources unless their estimated distances are poorly constrained. Then we examine the three sources part of the group of nine potential rotation-powered sources for which hard X-ray emission has been observed: SGR J1745--2900, 1E 1547.0--5408 and SGR 0501+4516. Fig.~\ref{fig:LxEdotHard} shows that 1E 1547.0--5408 remains still within the rotation-powered group while the other two sources do not. Thus, it becomes critical for these sources to verify the accuracy of the estimated distances and to explore the possible contribution of their associated supernova remnants to the hard X-ray emission.

\item
If these sources are powered by rotation, then other phenomena observed in known rotation-powered NSs could also been observed in these objects. Thus, we explored for the nine sources with $L_X <\dot{E}_{\rm rot}$ the possibility that the energetics of their bursting activity, $E_{\rm burst}$, can be explained from the rotational energy gained in an associated glitch, $\Delta E_{\rm rot}$. We thus computed lower limits to the fractional change of rotation period of NSs caused by glitches, $|\Delta P|/P$,  by requesting $\Delta E_{\rm rot} = E_{\rm burst}$. The fact that there exist physically plausible solutions for $|\Delta P|/P$ reinforces the possible rotation-powered nature for these sources (e.g., the cases of PSR J1846--0258 and PSR J1119--6127).
\item
We discuss in Sec.~\ref{sec:7} possible additional evidences pointing to the rotation-power nature of these nine sources. 1)  Radio emission is observed in four SGRs/AXPs and all of them are part of these nine sources. Radio emission characterizes ordinary pulsars but it is generally absent/unobserved in SGRs/AXPs. 2) We call also the attention to a peculiar emission property of the majority of these nine sources: seven of them belong to the group of the so-called transient sources (which are eleven in total). Within these seven transient rotation-powered objects we find the four showing radio emission. We argue that the analysis of the varying X-ray flux can given information on the NS properties and magnetospheric geometry, shedding light into the understanding of the properties of the radio emission. 3) Six of the nine sources have potential associations with supernova remnants, supporting a NS nature. See Table~\ref{tab9sources} for details.

\end{itemize}

Although we have shown the possibility that some SGRs and AXPs be rotation-powered pulsars, we are far from getting a final answer to the question of the nature of SGRs/AXPs. It is not yet clear whether all the current members of the SGR/AXP family actually form a separate class of objects, e.g. with respect to traditional pulsars, or if their current classification have led to misleading theoretical interpretations. Therefore we encourage further theoretical predictions and observations in additional bands of the electromagnetic spectrum such as the optical, high and ultra-high gamma-rays and cosmic-rays to discriminate amongst the different models and being able to elucidate the nature of SGRs and AXPs.

\section*{Acknowledgements}
It is a great pleasure to thank Sergio Campana for the comments and suggestions on the presentation of our results. Likewise, we would like to thank the anonymous referee for valuable comments.
JGC, RCR de Lima, and JAR, acknowledges the support by the International Cooperation Program CAPES-ICRANet financed by CAPES - Brazilian Federal Agency for Support and Evaluation of Graduate Education within the Ministry of Education of Brazil. MM and JGC acknowledge the support of FAPESP through the projects 2013/15088--0 and 2013/26258--4. JAR acknowledges partial support of the project No.~3101/GF4 IPC-11, and the target program F.0679 of the Ministry of Education and Science of the Republic of Kazakhstan.




\bibliographystyle{aa}
\bibliography{ref}
\end{document}